\documentclass{aa} 

\usepackage{txfonts}
\usepackage{graphicx}  
\usepackage{epstopdf} 
\usepackage{natbib} 
\bibpunct{(}{)}{;}{a}{}{,}

\begin{document}

\def\Rozanska{R\' o\. za\' nska~}
\def\Ka{K$_{\alpha}$}

\def\Ucolumndensity{cm$^{-2}$}
\def\Udensity{cm$^{-3}$}
\def\Uflux{erg~s$^{-1}$~cm$^{-2}$}
\def\Uxi{erg~s$^{-1}$~cm}
\def\Ua{erg$^2$~s$^{-2}$~cm$^{-4}$}
\def\Udistance{cm}
\def\Utemperature{K}

\def\diff{\mathrm{d}}

\def\titan{\textsc{TITAN}}


               
%
%
\title{Thermal instability in X-ray photoionized media \\
in Active Galactic Nuclei:}
\subtitle{II. Role of the thermal conduction in warm absorber}


\author{
B.~Czerny \inst{1} 
\and L.~Chevallier\inst{1,2}
\and A.C.~Gon\c{c}alves\inst{2,3,4}
\and A.~\Rozanska \inst{1}
\and A.-M.~Dumont\inst{2}
}
	 

\offprints{bcz@camk.edu.pl}

\institute{
Copernicus Astronomical Center, Bartycka 18, 
00-716 Warsaw, Poland 
\and
LUTH, Observatoire de Paris, CNRS, 
Universit\'e Paris Diderot, 5 Place Jules Janssen, 
92190 Meudon, France 
\and
Observatoire Astronomique de Strabsourg, Universit\'e 
Louis Pasteur, CNRS, 11 rue de l'Universit\'e, 
67000 Strasbourg, France
\and 
CAAUL, Observat\'orio Astron\'omico de Lisboa,
Tapada da Ajuda, 1349-018 Lisboa, Portugal 
}

\authorrunning{Czerny, Chevallier, Gon\c{c}alves et al.}

\titlerunning{Role of the thermal conduction }

\date{Received \today / Accepted (to be inserted later)}


\abstract
    {A photoionized gas under constant pressure can
display a thermal instability, with three or more solutions for
possible  thermal equilibrium. A unique solution of the structure 
of the irradiated medium is obtained only if electron conduction is considered.}
   {The subject of our study is to estimate how the efect of  
thermal conduction affects the structure and transmitted
spectrum of the warm absorber computed
by solving radiative transfer with the code {\sc titan}.}
   {We developed a new computational mode for the code {{\sc titan}} 
to obtain several solutions for a given external conditions and 
we test a posteriori which solution is the closest one to the 
required integral condition based on conduction.}
   {We demonstrate that the automatic mode of the code {{\sc titan}} 
provides the solution to the radiative transfer which is generally 
consistent with the estimated exact solution within a few per cent 
accuracy, with larger errors for some line intensities 
(up to 20 per cent) for iron lines at intermediate ionization state.}
   {}


\maketitle

%

\section{Introduction}

Recent modeling of the warm absorbing  medium in the central
regions of active galactic nuclei (AGN) has shown the 
 cloud models in constant total (gas plus radiation) pressure  
\citep{rozanska2006,goncalves2006,loic07} being more 
sophisticated with respect to those 
in constant density  \citep{netzer93,kaspi2001,kaastra2002},  
or models invoking a continuous 
wind \citep{steen2005,chelouche2005}. 
Constant pressure models, however, are not simple to deal with, as  
they require to address the subject of thermal instability. 

The existence of thermally unstable bubbles in gas affected by 
external illumination was
studied  by \citet{field65} with applications made to the interstellar 
and intergalactic medium.
More generally, thermal instabilities 
develop in photoionized gas in thermal equilibrium, i.e. in gas 
where the radiative heating is balanced by radiative cooling 
\citep[e.g.][]{krolik81}. 
These authors have shown that for a certain range of the ionization 
parameter, a medium illuminated by X-rays can achieve  local 
thermal equilibrium for three different values of the
temperature and the density.
Such multiple solutions provide a natural explanation for 
the two-phase gas scenario where colder and denser clouds at $\sim 10^4$ K are 
embedded in a hotter medium at $\sim 10^8$ K  and  
are at the origin of the description of the Interstellar Medium, 
the Warm Absorber and illuminated accretion disk atmospheres in terms of a  
clumpy  medium \citep{krolik81,smith88,raymond93,rozanska96,torri98,rozanska99, 
kriss2001}.

The uniqueness of the equilibrium state can be restored when the electron 
conduction is taken into account \citep[e.g.][]{begelman90,mckee90}. 
Mathematically, it happens because the 
energy balance equation with conduction becomes a differential instead of 
an algebraic equation.
Physically, the equilibrium distribution is achieved either by stationary  heat 
transport, or by evaporation or condensation of the material.

Analitycal estimations of the role of thermal conducion in planetary nebulae
\citep{SageSeaton73} have shown that thermal conduction is not an
important process for heating filamentary regions in whic the ionization is
maintained by diffuse radiation. 
 
The influence of thermal conduction in AGN  and galactic black holes (GBH) 
was studied in several papers, in the context of disk/corona boundary 
\citep{maciol97,rozanska99,dullemond99}, 
disk evaporation \citep{meyer94, liu99, rozanska2000b}, 
clumpy accretion inflow \citep{torri98}, 
and clumpy outflow \citep{krolik98}.
But the efect was never studied in the case of  warm absorbers, so
far. There is a big  importance to do it, since 
high resolutions spectra by X-ray statelites require proper 
models to deal with. 

Presently, no complex radiative transfer code for regions
illuminated by X-rays address the conduction term; therefore, 
codes tread thermal instabilities in two main approaches: i)
by fixing the discontinuity 
position at the optical depth found during the first integration
\citep{nayakshin2000}, ii) by finding a continuous intermediate  
but approximate solution, as discussed first by 
by \citet{rozanska02} and later in more details by 
\citet[][hereafter Paper I]{goncalves2007}  

In Paper~I we have  estimated the 
maximum error which can be made due to arbitrariness in the 
discontinuity location when the conduction is not included. 
In the present paper (hereafter Paper~II), we develop a  
method to determine the exact location of the discontinuity, 
by determining iteratively the warm absorber cloud structure  
and radiative transfer complete solution with a self-consistent 
location of the discontinuity and by comparing the so-obtained 
spectrum to the spectrum produced in a standard way by the 
\titan\ code. This position is crutial since in the models
of warm absorber under constant pressure the amount of energy
absrobed in lines strongly depends on the structure of 
the cloud. 

In the next section, we recall some generalities about the model 
and the \titan\ transfer-photoionization code, and we describe 
our approach to incorporate the effect of thermal conduction. Results are 
presented and compared to analitical model of \citet{SageSeaton73} in
Sect.~3.  Section~4 is dedicated to the physical 
discussion of these models.

\section{Methodology}
\label{sect:method}

It is well known that a numerical difficulty appears when trying to
estimate the importance of thermal conduction. The fundamental
quantity, i.e. the heat conductive flux, is proportional to the 
temperature gradient. In the numerical approach to the radiative 
transfer, the computations are performed using  a certain grid, 
defined by a given number of layers with a given thickness, that 
are used to reproduce the total gas slab.
This grid imposes a limit to the temperature gradient.  Even 
if the temperature distribution clearly shows a discontinuity, i.e. 
the temperature difference between two consecutive layers, $\Delta T$, 
is a large fraction of the highest value, $T$, the estimated 
temperature gradient, $\Delta T/\Delta z$ may not be large since the 
geometrical width of the layer, $\Delta z$, where the sharp 
temperature drop occurs, is set by the coarse spatial grid. In order 
to estimate accurately the heat conductive flux, the grid would 
have to be refined too much for practical use, since the actual 
spatial extension of the transition zone, is 
of order of the Field length \citep[see:][]{rozanska99}, 
and its optical depth for 
electron scattering, $\Delta \tau_{\rm es}$, can be estimated as 
\begin{equation}
\Delta \tau_{\rm es} = {\kappa_0 \over \Lambda}T^{7/4} 
\sigma_T = 5.3 \times 10^{-6} ~~T_6^{7/4} \Lambda_{-23}^{-1/2}, 
\end{equation}
where $\kappa_0$ is the conduction coefficient (see Appendix A), 
$\Lambda$ is the cooling function in $10^{23}$ erg cm$^3$ s$^{-1}$,
$\sigma_T$ - Thomson cross section, $T_{6}$ - temperature 
in $10^6$ K.   
Therefore, the direct approach which consists of including  
the conduction term into the computations can only be done in the 
case  of a very simplified approach to the radiative transfer, where
such a fine grid is acceptable \citep[e.g.][]{rozanska99}. 

In this study, we use an integral formulation of the solution to the
problem, which is not grid-dependent \citep{rozanska2000a}. The
method is based on the integral criterion for the location of the
discontinuity. Therefore, we do not include the conduction flux in the
energy balance but we aim at determining  the unique position of 
the discontinuity. To estimate this unique position, 
we solve the radiative transfer within an irradiated cloud under
constant pressure using the \titan\ code.

 \titan\ is a transfer-photoionization code developed by
 \citet{dumont2000,collin2004} to correctly 
model optically thick (Thomson optical depth up to several tens) 
ionized media; it can be applied equally to thinner media 
(Thomson depth $\sim$ 0.001 to 0.1).
The code solves full radiative transfer 
through the photoionized gas together with the ionization equilibrium, the  
thermal equilibrium and the statistical equilibrium of all the levels of
each ion. Our atomic data include $\sim 10^{3}$ lines from ions 
and atoms of H, He,  C, N, O,  Ne,  Mg, Si, S, and Fe. 
The code relies on 
the accelerated lambda iteration (ALI) method, which allows 
for the exact treatment of the transfer of both the continuum 
and the lines. It includes all relevant physical processes 
(e.g., photoionization, radiative and dielectronic 
recombination, fluorescence and Auger processes, collisional 
ionization, radiative and collisional excitation/de-excitation, 
etc.) and all induced processes. As an output, it 
gives  the ionization, density, and temperature structures, 
as well as the reflected, outward, and transmitted spectra. 
The energy balance is ensured locally with a precision of 
0.01\%, and globally with a precision of 1\%. For more details 
on the code and its evolution, see Paper~I and references 
therein. \\

The code presently offers  three operational modes: 
\begin{itemize}
\item[(i)] The first mode, used in a number of previous papers 
\citep[e.g.][]{dumont2002,rozanska02,collin2003}, is based 
on the subsequent (instead of simultaneous) determination of the
density and the temperature. As explained in Paper I, this solution 
is smooth (without a sharp discontinuity) but it does not satisfy accurately 
the condition of the constant pressure. Such a solution corresponds 
to the default, automatic mode of the \titan\ code and it will be 
denoted here  as an ``intermediate state'' (I).

\item[(ii)] The second mode, described in Paper I, preserves precisely the 
condition of the constant pressure and it locally finds multiple 
solutions to the temperature and the density in the instability 
zone. In this mode, the code is set to choose either the 
hottest solution, or  the coldest  solution, as the actual solution. 
In this way two separate solutions are obtained: one 
with the most extended hot zone at the cloud irradiated surface 
(state H) and the second  one with the narrowest hot zone (state
C).   

\item[(iii)] A third operation mode has been developed for the 
purpose of this work. In such a mode,   
multiple local solutions for the density and the temperature are 
searched for, but the actual discontinuous transition from the 
hot to the cold branch is set in advance, somewhere within the 
instability strip, at an arbitrarily imposed value of the 
column density, $N_H$. 
\end{itemize}

\subsection{Conduction flux and the location of the transition zone }
\label{sect:cond}

We have used this third operational mode of the code to 
calculate a series of models for an ionized cloud  
with a given set of parameters, only varying the  position 
$x$ of the temperature discontinuity. We denote such states as 
NH($x$), with the parameter $x$ being the column density in 
$10^{23}$ units, i.e. $x = N_H/10^{23}$~cm$^{-2}$.  For each of 
the models  NH($x$), we have calculated the value of the 
integral $A$ (right hand of Eq.~\ref{eq:balance-int2}, see Appendix A) 
with the purpose of verifying whether the position of the 
discontinuity is consistent with the criterion for the discontinuity 
position provided by the conduction flux ($A = 0.$; see 
Eq.~\ref{eq:balance-int}). In this way we estimated 
the position  $x$ for which the value of the integral $A$ is the 
closest to zero. The corresponding model was taken as the correct 
solution to the radiative transfer with the conduction term. We have 
then analyzed the properties of such a model, and we have compared 
them to the properties of the model obtained from the \titan\ 
computations in its automatic mode (state I). These results were 
used to estimate quantitatively the errors made in the transition 
spectra when the faster, automatic mode is used. This is important 
since the method used in the present paper to account for 
the electronic conduction, although physically correct, is  
highly time-consuming and it would be unrealistic to use as such 
for the interpretation of observational data.

\section{Results}

\subsection{Reference cloud}

Our reference cloud has been modeled  under 
constant total pressure equilibrium. It is characterized 
by an ionizing parameter $\xi=10^3$~\Uxi, a column density 
$N_H=2\times 10^{23}$~\Ucolumndensity, and a number density
$n_H=10^7$~\Udensity, at the illuminated side of the cloud. 
Such cloud parameters fall within the range of conditions valid 
for the Warm Absorber in active galactic nuclei. 

Abundances are cosmic (Allen 1973). This is a simple choice since the present work is aimed at testing the sensitivity of the cloud structure to the adopted description of the thermal instability. The actual abundanced in a given active galactic nucleus are unknown. Warm 
absorber material is outflowing from the nuclear region of an AGN, so this 
material is likely to have a long and complex history, being first an interstellar 
material of a host galaxy, later reaching the inner region where the 
intense starburst activity takes place (such activity frequently - or 
perhaps always - accompany the nuclear activity), the remnants of this 
starburst activity flow toward the black hole forming an accretion disk, 
and only next this material flows out in a form of disk wind before 
actually reaching a black hole. Therefore, in modelling a given active nucleus, we actually should 
determine the chemical composition from the fits to the data. In most 
cases, when such an attempt was made (mostly for iron, occasionally for 
lighter elements) the composition was rougly solar within a factor of a 
few. Therefore, the choice of standard composition is natural for the test purposes.

We have assumed this cloud to be  normally illuminated by a 
power-law incident flux with slope $\Gamma=2$ from  10 eV to 100 keV. 
Such a cloud is highly stratified in temperature, as it is subject 
to a thermal instability. In order to model our cloud, the medium 
has been divided into a few hundreds of layers, using an adaptive
grid. The first layer is hot ($T\sim 10^6 K$) and shows no 
thermal instability. The last layer is cold ($T\sim 10^4 K$) and 
it shows no thermal instability either. In the {innermost regions 
of the ionized cloud, two} instability zones exist (see Paper I), which 
may partially overlap. The whole thermal instability zone spans the 
range {\bf $N_H$} $\sim 1.65$--$1.92\times 10^{23}$~\Ucolumndensity.

\begin{figure}
\resizebox{\hsize}{!}{\includegraphics{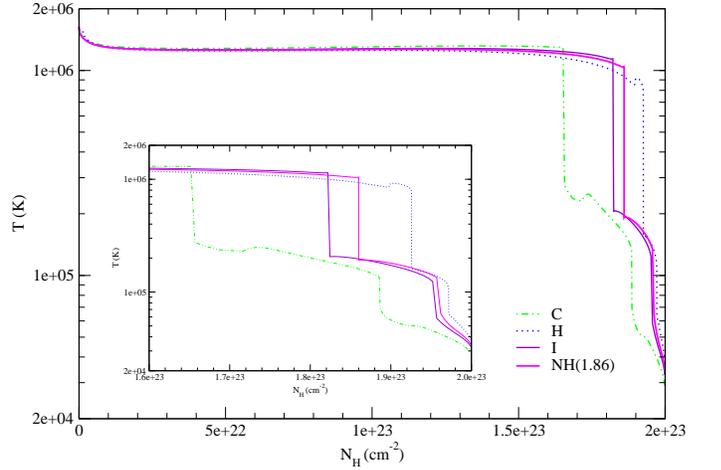}}
\caption{Temperature profile versus the column density for our 
  reference cloud in the extreme C and H states, in the intermediate I
  state resulting from \titan\  in its automatic operational mode, 
  and in the state NH(1.86) with the position of the temperature drop
  consistent with the conduction based criterion. {\it Inserted box:} 
  zoom of the temperature profile on the thermal instability zone.} 
\label{fig:temperature_solutions}
\end{figure}

The extension of the instability zones and consequently the limits 
to the exact location of the temperature discontinuity is provided 
by the two extreme models, H and C, corresponding to the lowest 
and the highest  values of the hydrogen column density measured 
at  the discontinuity.

For our reference cloud, the values of the hydrogen column are 
$1.92 \times 10^{23}$ cm$^{-2}$ for the state H, and 
$1.54 \times 10^{23}$ cm$^{-2}$ for the state C (in our notations, 
those values correspond to the NH(1.92) and NH(1.54) states, respectively).

The temperature profiles with the sharp drops for the states C and H
are shown in Fig.~\ref{fig:temperature_solutions}. 
Since the intermediate 
state is closer to the H state than to the C state, we will calculate 
the sequence of the intermediate states from NH(1.78) to NH(1.89).

\begin{figure}
\begin{center}
\resizebox{7.5cm}{!}{\includegraphics{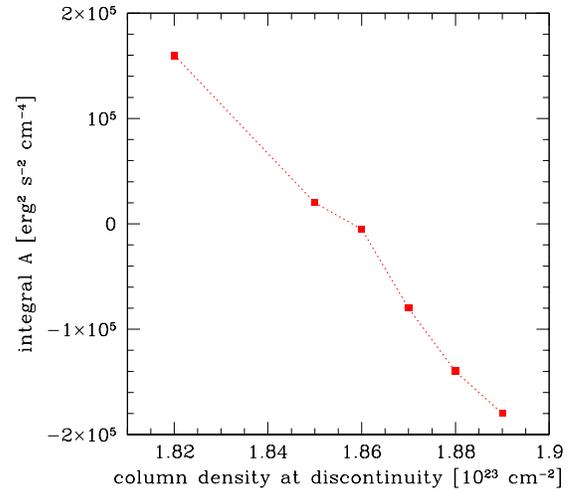}}
\end{center}
\caption{Variation of integral $A$ (units \Ua) as a function of the
column density (units \Ucolumndensity) for several intermediate states 
labeled after the adopted position of the discontinuity. The
best  agreement between the adopted position and the position
indicated by $A=0$ is obtained for model  NH(1.86).} 
\label{fig:hot_A}
\end{figure}

For all these states we have calculated the value of the
integral $A$ from Eq.~\ref{eq:balance-int} in the  Appendix A. 
This requires the construction of the heating/cooling curves as 
function of the temperature in various layers. The curves 
have a complex shape,  since multiple instability zones 
actually exist within the cloud. The integral gains a positive 
contribution from the beginning of the instability zone as the 
heating/cooling is positive, whereas, at the 
end of the transition region, the contribution is negative. 
This result is depicted in Fig.~\ref{fig:hot_A}.

The best agreement with the condition $A = 0$ is achieved for 
a given state NH($x$) with $1.85 < x < 1.87$, consistent 
with the real temperature drop being located somewhere between 
$1.85\times 10^{23}$~\Ucolumndensity and 
$1.87\times 10^{23}$~\Ucolumndensity. 
NH($x$) model computations being extremely time-consuming, 
we did not calculate the $1.85 < x < 1.87$ states very densely.   
Among the computed models, the NH(1.86) state offers the best 
agreement with the condition $A = 0$. 

Defining our equilibrium condition using
Eq.~\ref{eq:balance-int} given in the Appendix, we have 
assumed the left-hand part of this equation to be null, i.e. $A=0$. 
This requires that either $F_{cond}(T_{hot}) =
F_{cond}(T_{cold})$, or else that both $F_{cond}(T_{hot})$ 
and  $F_{cond}(T_{cold})$ are null (or at least negligible 
as compared to the integral, when it is non  null). We have tested 
that hypothesis {\it a posteriori}: calculating the  
left-hand side of Eq.~\ref{eq:balance-int2}, we have found it 
to be negligible.

Finally, it is interesting to note that the models  where the
adopted temperature drop roughly coincides with the estimated
location of the temperature drop from the conduction criterion   
are numerically stable (see Appendix B). 

In addition to the adopted NH(1.86) state, we also  
estimated the conduction flux above and below the temperature drop 
for a few additional cloud states. Our results show 
that the energy flux carried by electrons is always at least five 
orders of magnitude lower than the incident radiation flux. This is 
also true at the cloud surface since, although the temperature is 
higher there, the temperature drop is quite shallow. In an 
effort to further test our conclusions, we have refined the spacial 
grid by a factor of 100, at the expense of increasing the computational 
time considerably; our results show that the conduction flux
calculated directly from the temperature profile remain negligible. 
Therefore, we conclude that outside the instability zone, 
the electron conduction is negligible and the temperature drop is 
unresolved in \titan.  

\subsection{Transmission spectra}

In Paper~I we have shown that the transmission spectra for the H and 
C states are considerably different, thus demonstrating the 
importance of a sufficiently accurate measure of the temperature drop location. 
Thanks to the method developed in this paper, we were able to identify 
the (approximate) location that is consistent with the condition 
resulting from the introduction of the electron conduction term. We
are now capable to compare the transmission spectrum obtained for 
state NH(1.86) with the spectra resulting from states H and C. These 
were also compared with the standard I state resulting from the 
default \titan\ mode (faster and numerically more stable than the 
other modes).  

\begin{figure}
\resizebox{\hsize}{!}{\includegraphics{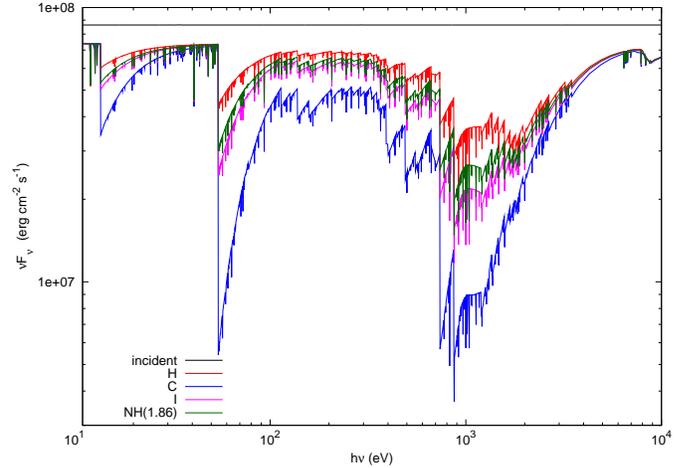}}
\caption{Transmitted continuum flux $\nu F(\nu)$ 
(in erg~s$^{-1}$~cm$^{-2}$~s$^{-1}$) vs. Energy (in eV) 
for our H, C, I, and NH(1.86) models. The plotted energy range 
is 10 eV to 10 keV; no differences are observed in the different
models outside this range. The spectral 
resolution $\nu/\Delta\nu$ is 300.}
\label{fig:transmitted}
\end{figure}

Figure~\ref{fig:transmitted} shows a pure transmitted spectra 
(i.e. covering factor $\sim 1$) at resolution $\nu/\Delta\nu=300$, 
and Table~\ref{tab:transmitted_lines} the comparison of the 
strongest lines present in  H, C, I, and NH(1.86) models. 
The continuum in NH(1.86) shows somewhat less absorption than 
in I, but the maximum difference, located around 1 keV, 
is never larger than 27\%. The maximum relative difference between 
the measured lines in NH(1.86) and I models is less than 24\% 
for the 200 strongest features. Most of the strongest lines 
show a difference less than 1\%, because they are saturated;  
some lines show a 20\% difference (e.g. \ion{Fe}{XII} and
\ion{Fe}{XIII}) which is related to a change of ionization fraction, 
due to a different ionic column density  (see Paper I). 
For the outward and reflected spectra (emitted in all directions 
{but the normal), the difference between models I and  
NH(1.86) is again weak, i.e. less than 25\%. We therefore 
conclude that} the I model can be used safely to describe such a
medium, which falls within the range of conditions valid 
for the Warm Absorber in AGN.

\begin{table}
\centering
\begin{tabular}{@{}lccl@{}}
\hline
ION & $h\nu$  & Line Flux       & $|$1-I/NH(1.86)$|$  \\
    &  (eV)   &    NH(1.86)     &  ~~~~~~(\%)  \\                 
\hline
     HI &  1.020e+01 &   -1.2e+05 &     ~~~~~~~1 \\
     HI &  1.209e+01 &   -7.2e+04 &     ~~~~~~~1 \\
   HeII &  4.080e+01 &   -7.0e+04 &     ~~~~~~~0.2 \\
   HeII &  4.836e+01 &   -5.7e+04 &     ~~~~~~~1 \\
     HI &  1.275e+01 &   -5.4e+04 &     ~~~~~~~0.2 \\
   HeII &  5.100e+01 &   -4.3e+04 &     ~~~~~~~1 \\
     HI &  1.305e+01 &   -4.0e+04 &     ~~~~~~~0.9 \\
 FeXVII &  8.260e+02 &   -3.9e+04 &     ~~~~~~~6 \\
   HeII &  5.223e+01 &   -3.6e+04 &     ~~~~~~~0.5 \\
  OVIII &  6.533e+02 &   -3.7e+04 &     ~~~~~~~6 \\
    CVI &  3.674e+02 &   -3.5e+04 &     ~~~~~~~5 \\
   FeXV &  4.366e+01 &   -2.7e+04 &     ~~~~~~~9 \\
  FeXXI &  1.001e+03 &   -3.4e+04 &     ~~~~~~~20 \\
 FeXXII &  1.042e+03 &   -3.3e+04 &     ~~~~~~~ 20 \\
   SiIX &  2.066e+02 &   -2.5e+04 &     ~~~~~~~7 \\
    SiX &  2.448e+02 &   -2.5e+04 &     ~~~~~~~5 \\
 FeXXII &  1.110e+02 &   -2.6e+04 &     ~~~~~~~3 \\
    SXI &  3.100e+02 &   -2.5e+04 &     ~~~~~~~4 \\
   SXIV &  2.908e+01 &   -2.5e+04 &     ~~~~~~~0.7 \\
   SiXI &  2.833e+02 &   -2.4e+04 &     ~~~~~~~5 \\
    CVI &  4.354e+02 &   -2.6e+04 &     ~~~~~~~7 \\
  SiXII &  2.450e+01 &   -2.4e+04 &     ~~~~~~~2 \\
     SX &  2.599e+02 &   -2.3e+04 &     ~~~~~~~6 \\
   FeXX &  9.336e+01 &   -2.5e+04 &     ~~~~~~~6 \\
FeXXIII &  9.334e+01 &   -2.4e+04 &     ~~~~~~~5 \\
  FeXXI &  1.213e+02 &   -2.3e+04 &     ~~~~~~~3 \\
   NVII &  5.001e+02 &   -2.4e+04 &     ~~~~~~~9 \\
  NeVII &  2.665e+01 &   -2.1e+04 &     ~~~~~~~6 \\
   OVII &  6.655e+02 &   -2.2e+04 &     ~~~~~~~0.05 \\
\hline
\end{tabular}
\caption{Table highlighting the differences between 
the transmitted spectra resulting from I and NH(1.86) models. 
For a small sample of 29 lines, we give the ion name, 
the energy of the line (eV), the line-integrated flux for 
  NH(1.86) (arbitrary units), and the relative difference 
with the I model (in percent). The lines are sorted from 
the strongest to the weakest, as observed in model NH(1.86).}  
\label{tab:transmitted_lines}
\end{table}

\subsection{The structure of the conductive layer}

\begin{figure}
\resizebox{\hsize}{!}{\includegraphics{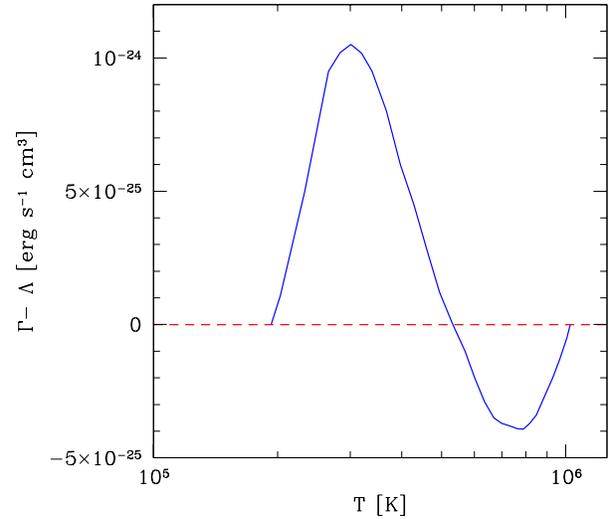}}
\caption{The heating/cooling curve for model NH(1.86) at the position of the 
conductive discontinuity as a function of the temperature, obtained under constant pressure condition. The curve crosses zero (radiative balance) for three values of the temperature. Compton heating as well as all atomic heating/cooling processes are included.}
\label{fig:curve}
\end{figure}

Since we determined the position of the transition layer we can calculate
{\it a posteriori} the structure of this sharp transition zone in a semi-analytical way, including the conduction flux. For that purpose we obtained first the heating/cooling function as a function of the temperature for the total pressure characterizing the model NH(1.86) in the transition zone. We stress that the heating/cooling is obtained for a constant presure, not constant density, as this assumption esentially influences its shape, as nicely explained in Goncalves et al. (2007). The resulting heating/cooling curve is shown in Fig.~\ref{fig:curve}.

\begin{figure}
\resizebox{\hsize}{!}{\includegraphics{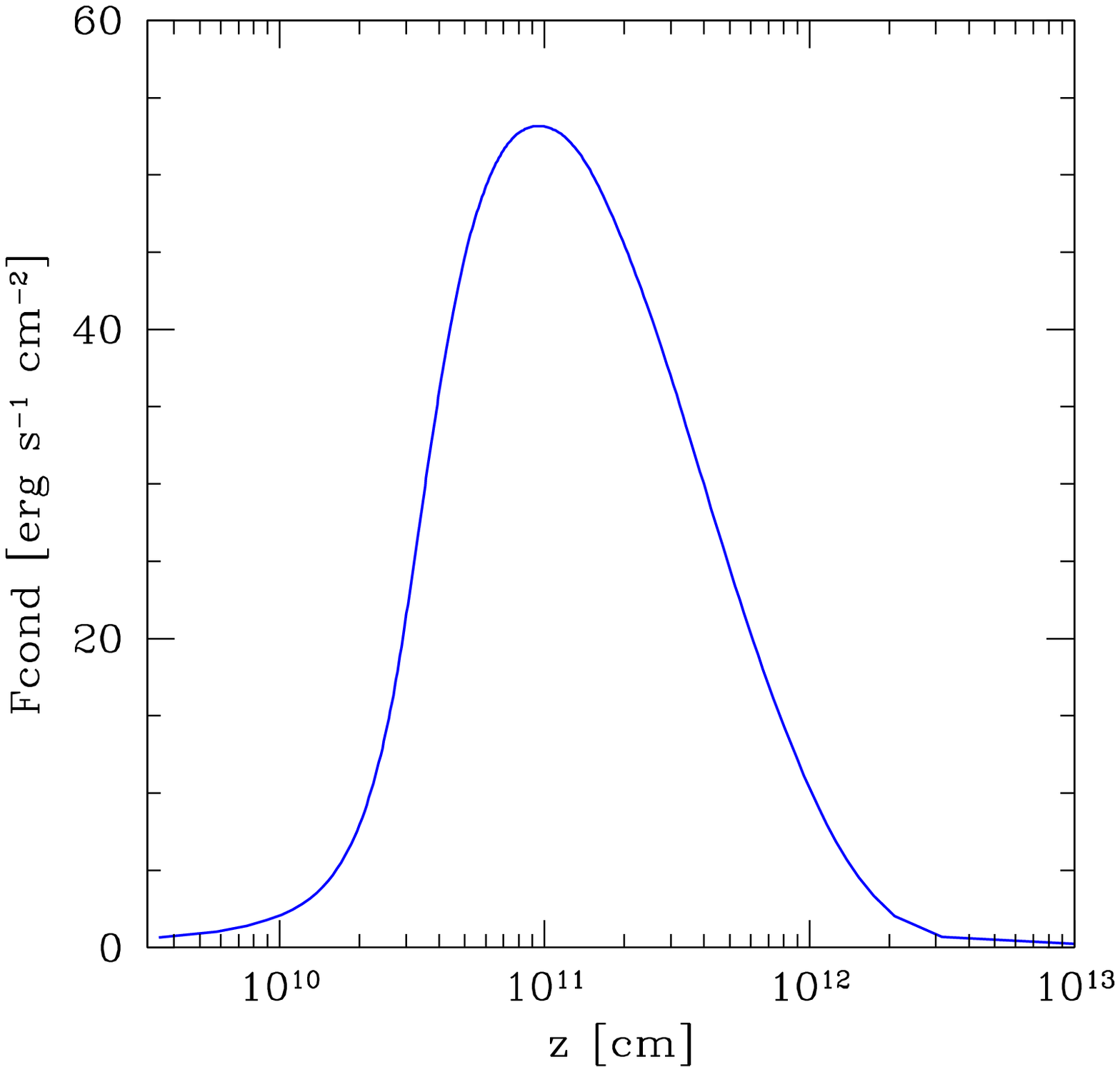}}
\resizebox{\hsize}{!}{\includegraphics{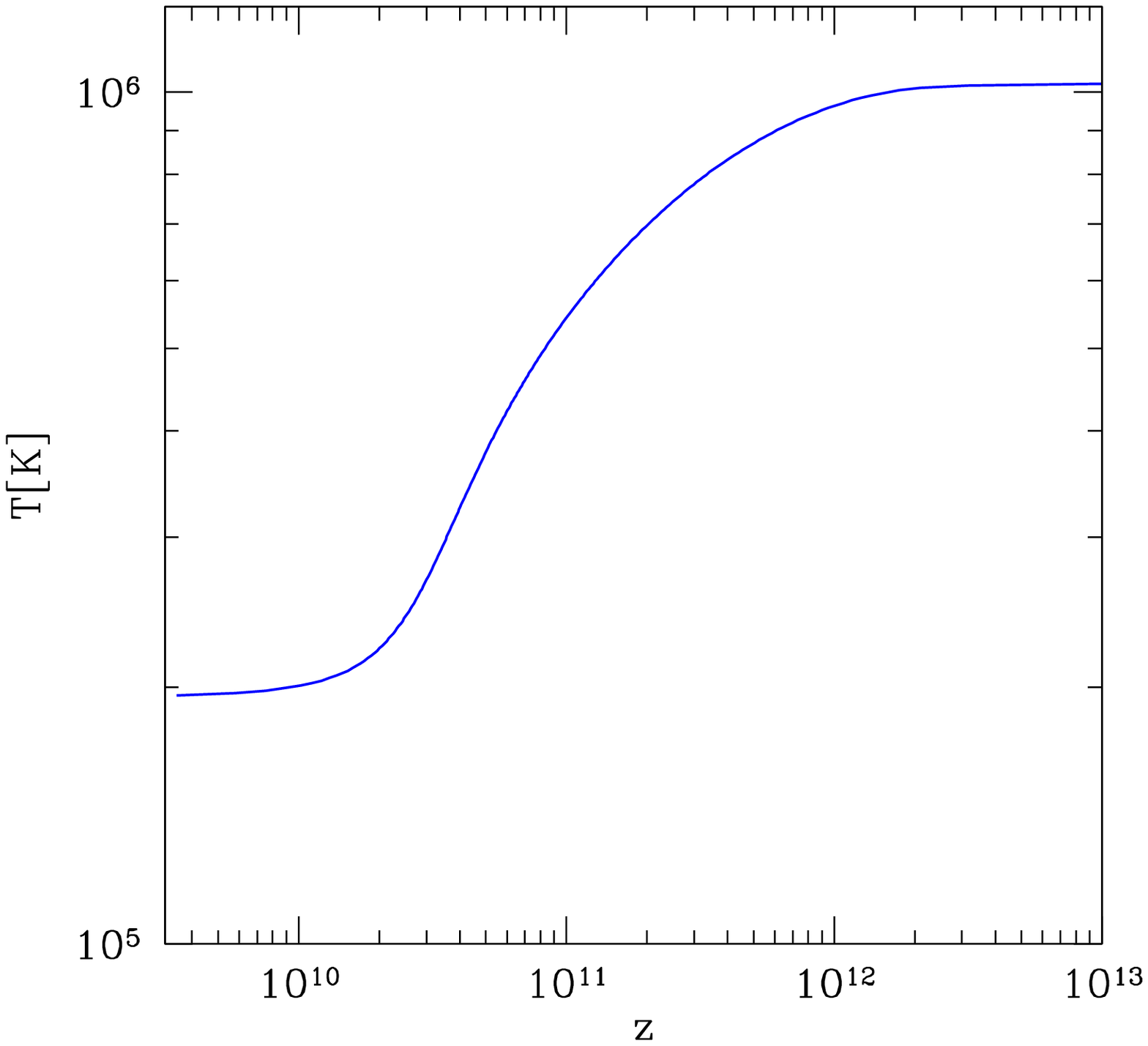}}
\caption{The spacially resolved temperature (upper panel) and conduction flux (lower panel) profiles in the transition zone obtained from semi-analytical approximation (see text). The transition is sharp in comparison with extension of a whole reference cloud $\sim 10^{15}$ cm.}
\label{fig:conduct}
\end{figure}

Having this curve, and assuming constant pressure in the sharp transition zone we can now easily obtain the conduction flux as a function of the temperature from the  integral form of the conduction equation (see Eq. 4 in the Appendix and replace $T_{hot}$) and replacing $T_{hot}$ with a set of intermediate value of $T$. Next, returning to the expression for conduction flux we can obtain the spacial temperature gradient in the constant pressure zone. The resulting flux and the temperature profile are shown in Fig.~\ref{fig:conduct}. The hydrogen column within this zone is equal to $1.54 \times 10^{20}$ cm so the zone unresolved in radiative transfer computations should not strongly contribute to the spectrum although it is not totally negligible.

In the transition zone the conductive flux is still orders of magnitude smaller than the radiative flux but nevertheless it contributes significantly to the total energy balace in this zone since the derivative of the conduction flux is large. The conductive flux balances the departure of the heating/cooling curve from zero and since the single heating or cooling mechanism gives values of order of $10^{-24}$ erg s$^{-1}$ cm$^3$ the flux derivative is of the same order of magnitude as the heat deposit or loss, just due to the geometrical narrowness of the zone.  Therefore, just comparing the two fluxes instead of their derivatives, as done e.g. by \citet{SageSeaton73} in their analysis may be misleading. 

The global significance of the conduction-dominated narrow zone itself of course is affected by its narrowness, and from this point of view the simple argument of \citet{SageSeaton73} is correct and the zone does not contribute strongly to the final transmission spectra so the fact that in our numerical computations of the radiative transfer we neglected this zone completely is fully justified. 

The important effect of electron conduction is, however, in the determination of the position of the discontinuity, and this is the real effect we search for, as explained in Sect.~\ref{sect:method}. Having semi-analytical results we can now check if the conduction is effective enough to move the position of the transition zone.

If the initial position of the transition zone for some reason is not consistent with the conduction criterion (see Appendix) the transition zone will move with the velocity rougly given by the ratio of the conduction flux to the local energy density, i.e. $2 F_{cond}/(5n_HkT) \sim 10$ km s$^{-1}$. This speed is strongly subsonic. The whole size of the reference cloud where three solutions for the radiative equilibrium are possible covers the range of $3 \times 10^{22}$ cm$^{-2}$ in hydrogen colun and the spatial distance of $\sim 10^{14}$ cm, i.e. the conduction flux will adjust the front position in about 3 years. Slowly moving ($v < 500$ km s$^{-1}$) distant (above $\sim 3 \times 10^{16}$ cm) cloud in the radiation field of an active nucleus has time to reach such equilibrium since the motion of the front due to systematically decreasing flux is slower than the effect of conduction estimated above. The fact that the irradiation in an AGN is time-dependent complicates the problem, as discussed by Chevallier et al. (2007).

A question remains whether such a narrow zone can exist and be stable. The first issue is related to the mean free path of electrons. In the likely presence of a weak magnetic field this seems to be possible. The mean free path of an electron for a micro-Gauss magnetic field is a fraction of a cm, much smaller than the size of the transition region. Stability issue is another problem. A study e.g. by \citet{rozanska99} in the context of disk/corona border indicated marginal stability. 

However, a magnetic field may also reduce heat conduction and in this case we may strongly overestimate the conduction flux. The issue was discussed other astrophysical contexts like wind collisions but the theory is rather complex and uncertain. Recent Chandra observations of a planetary nebule PN BD+30$^{\circ}$3639 by \citet{nordon2009} suggest complex structure of such a zone, with ions penetrating the contact discontinuity boundary. The countrate in AGN observation is much lower so at present the search for such traces in AGN warm absorber is likely to be premature.

\section{Discussion}

This study represents an important contribution to the understanding 
of the technical aspects behind radiative transfer computations 
in stratified media under total constant pressure. It is well 
known that an irradiated gas in constant pressure will develop 
thermal instabilities, if the electron conduction is neglected 
\citep[e.g.][]{krolik81}.  These instabilities 
appear as technical problems in radiative transfer and 
photoionization codes. Therefore, various codes use different 
approaches to address this problem.

In this paper we tested the different numerical approaches 
used by the \titan\ code. We constructed a subset of radiative 
transfer solutions, all for the same column density of the cloud 
and the same incident radiation flux, and identified the one 
which is close to satisfy the equilibrium condition derived from the 
electron conduction theory. We found that the approximate 
solution obtained by the standard, automatic mode of the 
\titan\ code provides a satisfactory representation of 
the transmitted spectra. 
We find that most line intensities resulting from 
this automatic mode are in agreement with the values derived 
from the electron conduction theory, reaching an accuracy of the 
order of 1\%,  with only a few 
exceptions (mostly iron lines at intermediate ionization state) 
where the discrepancy between the solutions reached 20\%. 

We have tested our methodology and hypothesis on a given 
cloud, described by a single set of  parameters; these were 
chosen among the range of conditions usually met in the modeling 
of the X-ray ionized gas near the central regions of AGN, and 
are fully compatible with a Warm Absorber medium.  An issue 
remains --  that of computations based on the conduction 
criterion being much more time-consuming (at least by a 
factor 20), than the standard radiative transfer computations, 
which are already quite long. Therefore, it is not feasible to 
apply this computation method to directly obtain solutions for 
a grid of models, for instance.  The results of this work 
provide, nevertheless, important evidence that the standard,  
automatic computing mode of the \titan\ code  is not 
strongly affected by systematical errors caused by neglecting 
to include the electron conduction. This is  important, since 
constant pressure models are a much better approximation than 
constant density models, with applications in several physical 
circumstances, e.g. the Warm Absorber clouds, or the disk/corona 
transition zone in an accretion disk onto a black hole. 

On the other hand, our analysis does not address the evolutionary
aspect of the presence of thermal instability. 
In this paper, as in other papers devoted to the modeling of 
the Warm Absorber, we assumed that the very existence of the 
clouds (instead of a continuous wind) is a result of this 
instability, but we did not address the timescales for the 
cloud formation and its survival. A first attempt at this exercise 
can be found in Paper~I and in \citet{loic07}. 
Additional discussions on these aspects of the thermal instability 
have been reported in other contexts i.e.: the hot medium in galaxy
clusters \citep{kim2003}, and the multiphase interstellar medium
\citep{inoue2006}. 
In the case of AGN the issue of the medium fragmentation is
additionally made more complex by the fact that the irradiation flux 
itself varies significantly in time, so the cloud can never achieve 
a full thermal equilibrium \citep[e.g.][]{loic07}. 
Those issues are clearly beyond of the scope of the present paper.

\begin{acknowledgements}
We thank  S.~Collin and V.~Karas for helpful discussions, 
and we are grateful for the hospitality provided by 
the Astronomical Institute, Academy of Sciences of the Czech Republic, 
Prague, where part of this project was completed.   
A.~C. Gon\c{c}alves acknowledges support from the {\it
Funda\c{c}\~ao para a Ci\^encia e a Tecnologia},  Portugal, 
under grant no. BPD/11641/2002. 
Part of this work was carried out within the framework of the European
Associated Laboratory "Astrophysics Poland-France", and was supported
by grant 1P03D00829 of the Ministry of Science and Education, and the 
Polish Astroparticle Network 621/E-78/BWSN-0068/2008.
\end{acknowledgements}

\bibliographystyle{aa}
\bibliography{refs}

\section*{Appendix A}

We recall the main relations for energy balance with the thermal
conduction effect included \citep[see][]{rozanska2000a}, which are
used and eventually modified for the purpose of this  study.
The energy balance at $z$ (i.e. which permits to compute the
temperature at this position) across the boundary between the hot and
cold media in the static case is given  by 
\begin{equation}
\label{eq:balance-diff}
\frac{\diff F_{cond}}{\diff z} = n_e n_H ( Q^+ - \Lambda )\;,
\end{equation}
where $Q^+$ and $\Lambda$ determine the heating and cooling rates of
the matter by energy exchange with radiation only, respectively, and
$n_e \sim 1.18 n_H$ for such cosmic abundances. Note that the total
density is $n = 2.25 n_H$. Note also that $Q^+$ and $\Lambda$ are the
usual definition of volumic coefficients par particle, explaining the
factor $n_e n_H$. On the left-hand side of Eq.~\ref{eq:balance-diff}
appears  the conductive heat flux $F_{cond}$, defined as
\begin{equation}
\label{eq:fcond}
F_{cond} = - \kappa_0 T^{5/2} \frac{\diff T}{\diff z}\;,
\end{equation}
where $\kappa_0 = 5.6\times 10^{-7}$~erg~cm$^{-1}$~s$^{-1}$~K$^{-7/2}$
is  the conductivity constant for a typical astrophysical plasma 
with cosmic abundances \citep{allen73}. The typical value of
$\Lambda $ is of order of $ 10^{-23}$~erg~cm$^3$~s$^{-1}$. In the  code
Titan, the term on left hand side of Eq.~\ref{eq:balance-diff} is 
neglected while the right hand side is calculated carefully, with 
line and continuum emission/absorption processes.

Under the condition of constant gaseous pressure (i.e. perfect gas),
we can express Eq.~\ref{eq:balance-diff} performing an integral 
(Eq.~8 in \citet{rozanska2000a}) over the temperature in the 
thermal instability region from the
cold solution $T_{cold}$ to the hot solution $T_{hot}$. Under the
constant total (gaseous + radiative) equilibrium of our clouds
$P_{tot} = P_{gas} + P_{rad}$, both gaseous and radiative pressures
vary in the thermal instability region, in a non-local dependence of
the temperature. This assumption cannot be done on theoretical
grounds. However we observe that this variation is weak, so the
non-local dependence is not drastic, and use the local value of the
gaseous and radiative pressure is consistent with the way we have
computed the heating and cooling curves for our code. Eq.~8 from
\citet{rozanska2000a} becomes, when inserting the factor $n_e n_H
= 1.118/(2.25 k)^2 (P_{tot}-P_{rad})^2 / T^2$,  where $k=1.38\times
10^{-16}$~erg~K$^{-1}$  is the Boltzmann constant,
\begin{eqnarray}
\label{eq:balance-int2}
\lefteqn{ F_{cond}^2 (T_{cold}) - F_{cond}^2 (T_{hot}) =} && \nonumber \\
&& 2 \kappa_0 \frac{1.18}{2.25^2 k^2} (P_{tot} - P_{rad})^2 
\int_{T_{cold}}^{T_{hot}} (Q^+ - \Lambda) \sqrt{T} \diff T \,,
\end{eqnarray}
where the conductive heat flux at temperatures $T_{hot}$ and
$T_{cold}$ should be small enough to be neglected. We denote by $A$
the right-hand part of Eq.~\ref{eq:balance-int}. Usual calculation of
the thermal conduction effect require $A=0$. Each quantity $F_{cond}^2
(T_{hot})$ or $F_{cond}^2 (T_{cold})$ should be small, as the
temperature gradient outside the thermal instability zone is
small. But the difference could be of the same order of magnitude than
$A$. We will  neglect the left-hand side of
Eq.~\ref{eq:balance-int2} and we will check the consistency of the 
assumption {\it a posteriori}: 
\begin{equation}
\label{eq:balance-int}
2 \kappa_0 {{1.18} \over {2.25^2 k^2}} (P_{tot} - P_{rad})^2 
\int_{T_{cold}}^{T_{hot}} (Q^+ - \Lambda) \sqrt{T} \diff T  = A = 0.
\end{equation}
%

\section*{Appendix B}

\begin{figure}
\resizebox{\hsize}{!}{\includegraphics{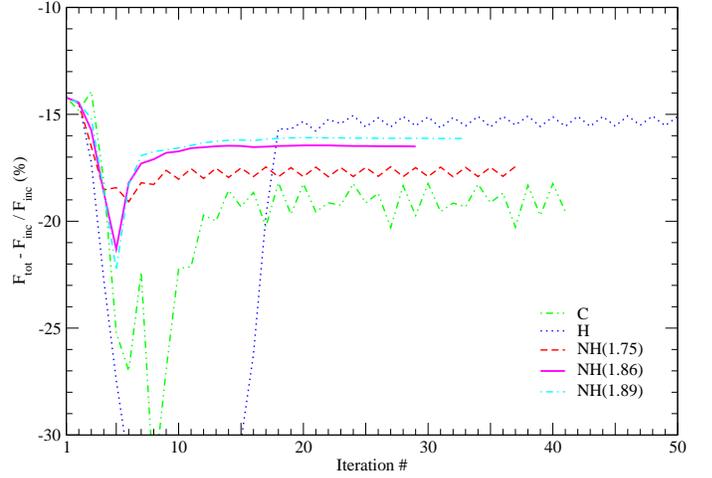}}
\caption{The convergence properties of our reference cloud as a 
function of the iteration number up to iteration 50, for the 
states H (no change up to iteration 65)  C,  and NH(1.86). Note 
that the NH(1.86) model converges more rapidly and without 
oscillations.}  
\label{fig:bil_hot}
\end{figure}

It is interesting to note that finding the exact location of the 
discontinuity in the model provides also a solution to some 
numerical problems which are met in {\sc titan} used in the second 
(nonstandard) operational mode, used in Paper I. 
If we attempt to determine the type H state of the cloud (i.e. we
choose always the hottest solution in the multiple solution region;
see Sect.~\ref{sect:method}) we actually face the problem of the small regular
oscillations in the state properties, which translate into the 
small but finite oscillations in the total energy balance 
of the cloud. To illustrate this we present in
Fig~\ref{fig:bil_hot} the dependence of our estimator
of convergence, on the number of iterations for different states. 
The oscillations in total
energy balance correspond to the oscillations in the location of the
main temperature drop (see Fig.~\ref{fig:temperature_solutions}). 

The oscillations cease to exist if we impose the position of the 
temperature discontinuity quite close to the position required by 
the conduction condition $A = 0.$. Specifically, oscillations are 
not present in all models from NH(1.82) to NH(1.89).

\end{document}